\documentclass[twocolumn]{aastex631}

\shorttitle{Gas shielding of soft binaries}
\shortauthors{Rozner \& Perets}

\usepackage{graphicx}	
\usepackage{amsmath}	
\usepackage{amssymb}	
\usepackage{color}
\usepackage{float}
\usepackage{ulem}
\usepackage{soul}
\usepackage{lipsum}
\usepackage[T1]{fontenc}

\usepackage{xcolor}

\DeclareRobustCommand{\VAN}[3]{#2}
\let\VANthebibliography\thebibliography
\def\thebibliography{\DeclareRobustCommand{\VAN}[3]{##3}\VANthebibliography}

\usepackage{graphicx}	
\usepackage{amsmath}	
\usepackage{braket}

\begin{document}

\title{Soft no more: gas shielding protects soft binaries from disruption in gas-rich environments}


\email{morozner@campus.technion.ac.il}

\author[0000-0002-2728-0132]{Mor Rozner}
\affiliation{Technion - Israel Institute of Technology, Haifa, 3200002, Israel}


\author[0000-0002-5004-199X]{Hagai B. Perets}
\affiliation{Technion - Israel Institute of Technology, Haifa, 3200002, Israel}
\affiliation{Department of Natural Sciences, The Open University of Israel, 1 University Road, PO Box 808, Raanana 4353701, Israel}

\begin{abstract} 
Binaries in dense environments are traditionally classified as soft or hard based on their binding energy relative to the kinetic energy of surrounding stars. Heggie's law suggests that stellar encounters tend to soften soft binaries and harden hard binaries, altering their separations. However, interactions with gas in such environments can significantly modify this behavior. This study investigates the impact of gas on binary softening and its consequences.
We find that gas interactions can actually harden binaries, extending the soft-hard boundary to larger separations. This introduces a "shielding radius" within which binaries are likely to harden due to gas interactions, surpassing the traditional soft-hard limit. Consequently, a notable portion of binaries initially classified as "soft" may become "hard" when both gas and stars are considered. We propose a two-stage formation process for hard binaries: initial soft binary formation, either dynamically or through gas-assisted capture, followed by gas-induced hardening before eventual disruption.
In environments with low gas density but high gas content, the shielding radius could exceed the typical hard-soft limit by an order of magnitude, leading to a significant fraction of originally soft binaries effectively becoming hard. Conversely, in high gas-density environments, gas-induced hardening may dominate, potentially rendering the entire binary population hard.
Gas hardening emerges as a crucial factor in shaping binary populations in gas-rich settings, such as clusters, star-forming regions, and possibly AGN disks. This highlights the complex interplay between gas dynamics and stellar interactions in binary evolution within dense environments.  
\end{abstract}



\section{Introduction}
Binaries in dense environments could be categorized into two groups, based on their energy relative to the mean energy of their environment. Soft binaries are defined as binaries with low energies relative to the mean energy of binaries in the cluster, i.e. $|\tilde E|\lesssim \braket{m\sigma^2}$, while hard binaries have energies that exceed the mean energy of the cluster, i.e. $|\tilde E|\gtrsim \braket{m\sigma^2}$. 
The evolution of the two groups differs qualitatively from each other. 
Heggie's law states that soft binaries tend on average to get softer, while hard binaries tend to get harder \citep{Heggie1975,Hills1975}, albeit this could be somewhat modified when accounting for external potentials \citep{GinatPerets2021}.

The analysis of hard/soft dynamics has considered only dynamic interactions between stars in dense star clusters. However, under a wide range of conditions, dense stellar clusters can be highly enriched in ambient gas. This naturally occurs in the early star-forming epoch, at which stars are formed from the collapsing and fragmenting gaseous background and are embedded in gas until the gas dissipates and/or accreted. Furthermore, older clusters can also be enriched in gas, e.g. stars embedded in AGN (active galactic nucleus) disks, nuclear clusters enriched in gas from infalling gas clouds, and/or gas ejected by evolved stars through winds. Similarly, globular clusters, in which multiple populations of stars are observed suggest multiple star-formation epochs (e.g. \citealp{Carretta2009,Renzini2015,Gratton2019,BastianLardo2018}), in which earlier-formed stars from the previous generation(s) would be embedded in newly enriched gas needed for the formation of a later generation of gas.

In such gas-enriched environments, in addition to stellar interactions, the interaction with the gas leads to energy dissipation and further unique dynamical phenomena.
Accretion onto stars/compact objects in gaseous environments could give rise to X-ray flares \citep{BahcallOstriker1976} or
explode via supernovae \citep{Ostriker1983,Artymowicz1993}. 
Gas hardening was investigated thoroughly,
in molecular clouds \citep{Stahler2010}, protoplanetary disks \citep{PeretsMurrayClay2011,GrishinPerets2016},
as a catalyzer for various astrophysical processes including gravitational waves (GWs) mergers \citep{McKernan2012,Stone2017,Tagawa2020,
RoznerGW2023}, gas-assisted binary formation \citep{Tagawa2020,capture2023,
Li_etal_2023,Rowan2023,Whitehead2023,LiLai2024} and enhanced formation of massive black holes \citep{RoupasKazanas2019}.

Binary formation in gas could be divided into three categories (e.g. \citealp{Bonnell2001,Kratter2011} 
and references therein): in-situ formation in which the binaries are formed following the collapse of gas clumps in the separation in which they are observed today (which includes fission and fragmentation), collapse formation in a wider separation than their current, followed by a migration triggered by a dissipative force;  
and gas-assisted binary capture \citep{Tagawa2020,RoznerGW2023,Whitehead2023}.
Some of the suggested dissipation mechanisms following the formation take place either through an interaction with a circumstellar disk \citep{ClarkePringle1991,HallClarkePringle1996}, or through interaction with the ambient surrounding gas 
Other binary-formation channels include 3-body interaction 
(\citealp{Mansbach1970,AarsethHeggie1976,GoodmanHut1993,GinatPerets2024}), tidal capture \citep{PressTeukolsky1977}, dynamical friction assisted capture, also known as L2 \citep{GoldreichLithwickSari2002}, and gas-drag formation of planetesimal binaries \citep{Ormel2010,PeretsMurrayClay2011}. 

Although some of these mechanisms occur in naturally gas-rich environments and are also dense with stars, such as AGN disks, the hardness/softness of binaries, or the interconnection between interactions induced by the stellar and gaseous background, has been little explored in this context. Here we focus on these issues.

We examine the effect of gas-hardening on binaries in GCs, together with stellar hardening/softening. We derive analytically the new soft-hard boundary, accounting for the effect of the gas. We then discuss further implications on cluster dynamics and the population of hard/soft binaries. 

In section \ref{sec:dynamics in gaseous environments}, we briefly overview the components that different contributions to the dynamics of binaries in gas. In section \ref{sec:gas shielding} we introduce the concept of gas shielding. In section \ref{sec:discussion} we discuss our results, possible caveats and future implications. In section \ref{sec:summary}, we summarize our results and conclude.  

\section{Dynamics of binaries in gaseous environments}\label{sec:dynamics in gaseous environments}

There are several approaches to model dynamics in gas. Some of them are mini-accretion disks \citep{Artymowicz1991,McKernan2012,Stone2017,Tagawa2020}, simulating the Bondi-Hoyle accretion supersonic flows and deriving the corresponding energy dissipation rate \citep{Antoni2019} 
and gas dynamical friction \citep{Ostriker1999}.   
 In this paper, we will use the latter unless stated otherwise.
While the details differ from one model to another, all of them model energy dissipation, and hence the concept of the process we present below would also hold for other descriptions of motion in gas, with the proper modifications.  
The motion of objects in gas is still not completely understood and is under active research, with some fundamental issues still debated, including the direction of the migration (e.g. \citealp{Moody2019,DongMunoz,Grishin2023,Duffell2024_comparison}). 

\subsection{Hardening/softening through dynamical encounters}

A binary is called hard if its energy exceeds $\bar m \sigma^2$. This condition sets for every pair of masses a critical semimajor axis. Binaries with larger semimajor axes will be soft and binaries with smaller semimajor axes will be hard. 

\begin{align}\label{eq:aSH}
a_{\rm SH}= \frac{Gm_1m_2}{2\bar m \sigma^2}
\end{align}

Binary softening is dominated by a series of distant encounters, gradually increasing the internal energy of the binary, and could potentially lead to positive energy, i.e. disruption of the binary. The softening rate could be calculated using the diffusion coefficients \citep{Heggie1975,BinneyTremaine2008}, 

\begin{align}
\braket{\dot E_{\rm soft}}\approx \braket{D[\Delta \tilde E]}\approx  \frac{8\sqrt{\pi}G^2\mu \bar m \rho_\star \ln \Lambda_{\rm bin}}{\sigma}
\end{align}

\noindent
where the Coulomb factor is $\Lambda_{\rm bin}= a\sigma^2/(4G\bar m)$.

The rate at which hard binaries become harder 
is \citep{Heggie1975,Spitzer1987,
HeggieHut1993, BinneyTremaine2008,Celoria2018}

\begin{align}
\braket{\dot E_{\rm hard}}=
2\pi \frac{G^2 m_1m_2 \rho_\star (M_{\rm bin}+\bar m)}{M_{\rm bin}\sigma}
\end{align}

\noindent
Up to a factor of order of unity, where $E_{\rm bin}$ is the energy of the binary and $\bar m$ is the (mean) mass of the perturber.  
For an analytic derivation of the hardening rate considering both energy and angular momentum in the equal masses case see \citealp{GinatPerets2021_3body}.

The softening/hardening rates in terms of $a$
are given correspondingly by 

\begin{align}
&\frac{da}{dt}\bigg|_{\rm soft}=\frac{16\sqrt{\pi}G\bar m \rho_\star \ln \Lambda_{\rm bin}}{M_{\rm bin}\sigma}a^2, \\
&\frac{da}{dt}\bigg|_{\rm hard}=
-2\pi \frac{G\rho_\star(M_{\rm bin}+\bar m)}{\sigma M_{\rm bin}}a^2
\end{align}

\noindent
where $\rho_\star$ is the background density of the stellar perturbers.
It should be noted that while binary hardening/softening is usually considered relative to a background of $\bar m$, in some cases, less frequent interactions with more massive objects could lead overall to more significant effects. Black holes in the background of other black holes are considered hard for semimajor axes lower than $a_{\rm SH, \bullet}\approx 1.35 \rm{AU}$ (see \citealp{Quinlan1996,KritosCholis2020,RoznerGW2023}), while for stellar background, they are considered hard for semimajor axes below $a_{\rm SH, \star}\approx 200.53 \ \rm{AU}$. In between, these binaries are considered hard relative to other black holes but soft relative to stars. 
However, the contribution from stellar hardening dominates over the softening due to black holes, such that we can define all the stellar softening/hardening based on stellar background only. 

\begin{align}
&\bigg|\frac{da/dt|_{\rm soft,\bullet}}{da/dt|_{\rm hard,\star}}\bigg|= 
\frac{8}{\sqrt{\pi}}\frac{M_\bullet}{M_{\rm bin}+m_\star}\frac{\rho_\bullet}{\rho_\star}\ln \Lambda_{\rm bin}\sim10^{-2}\ll 1
\end{align}

\subsection{Gas dynamical friction}

The GDF force on an object with mass $m$ is \citep{Ostriker1999},

\begin{align}\label{eq:FGDF}
\textbf{F}_{\rm GDF}=-\frac{4\pi G^2m^2 \rho_g}{v_{\rm rel}^3}\boldsymbol{v}_{\rm rel}
I(v/c_s)
\end{align}

\noindent
where $G$ is the gravitational constant, $\rho_g$ is the gas density, $c_s$ is the sound speed, and $\mathbf{v_{\rm rel}}$ is the relative velocity between the object and the gas. The function $I$ is given by 

\begin{align}
I(\mathcal M)=
    \begin{cases}
        \frac{1}{2} \log(1-\mathcal M^{-2})+\ln \Lambda, & \mathcal M> 1\\
        \frac{1}{2} \log\left(\frac{1+\mathcal M}{1-\mathcal M}\right)-\mathcal M, & \mathcal M<1\\
    \end{cases}
    \label{eq:I}
\end{align}

\noindent
where $\mathcal M = v/c_s$ is the Mach number. 

The semimajor axis evolution is given by

\begin{equation}
\begin{aligned}\label{eq:a_evolution}
&\frac{da}{dt}\bigg|_{\rm GDF}= - \frac{8\pi G^{3/2}a^{3/2}}{\sqrt{m_1+m_2}}\rho_g(t)\frac{m_1}{v_{\rm rel}^2}I \left(\frac{v_{\rm rel}}{c_s}\right)\xi(q), \\ 
&\xi(q)=(1+q^{-1})^2+q(1+q)^2
\end{aligned}
\end{equation}

\noindent
where $q=m_2/m_1$ is the mass ratio of the binary.
Note that this equation differs from eq. 7 in \cite{RoznerGW2023} by a factor of unity, due to mass-ratio corrections. The relative velocity is taken as $\max\{\sigma,v_{\rm Kep}\}$. The energy evolution is given by

\begin{align}
\frac{dE}{dt}\bigg|_{\rm GDF}=-\frac{4\pi G^{5/2}m_1^2m_2}{\sqrt{m_1+m_2}}\frac{\rho_g}{v_{\rm rel}^2}I\left(\frac{v_{\rm rel}}{c_s}\right)\xi(q)a^{-1/2}
\end{align}

\section{Gas shielding}\label{sec:gas shielding}

The process of hardening/softening binaries could significantly change in the presence of gas. The interaction with gas leads to energy dissipation, which could potentially compete with the stellar softening of soft binaries. Binaries that were considered soft, accounting for stellar interactions only, could become hard when adding the contribution from gas-hardening.  We term this process as \textit{gas shielding}, as the gas shields soft binaries from disruption, by dissipating their orbits, making them hard binaries (with respect to stellar encounters) before encounters with other stars soften and eventually disrupt them. Hence, the definition of soft/hard binaries should be revised in the presence of gas, to include the contribution of gas hardening. 
Hereafter, we will demonstrate the effect of gas-shielding in GCs (and briefly consider also other gas-rich environments). This is a general process, that could occur in general in any gas-rich media, with the proper modifications. 

For any set of parameters of the binary and its environments, one can define the \textit{shielding radius}. The shielding radius is the critical semimajor axis between soft and hard binaries, i.e. the hardest soft binary/the softest hard binary, when we consider the combined effect of stellar hardening/softening and gas hardening. 

For a given gas density, stellar number density, and typical velocity dispersion, the shielding semimajor axis could be calculated by setting $\dot a_{\rm tot}=\dot a_{GDF}+\dot a_{\rm soft}=0$, which is given by  

\small
\begin{align}\label{eq:a_crit}
a_{\rm{shield}}^{1/2}\ln\left(\frac{a_{\rm shield}\sigma^2}{4G\bar m}\right)= \frac{\sqrt{\pi}}{2}\frac{\sqrt{GM_{\rm bin}}}{\sigma}\frac{\rho_g}{\rho_\star}\frac{m_1}{\bar m}I\left(\frac{\sigma}{c_s}\right)\xi(q)
\end{align}
\normalsize

which yields an analytical solution 

\begin{align}\label{eq:rho_crit}
&a_{\rm shield}= \frac{B^2}{4W^2\left(-\frac{B\sigma}{4\sqrt{G\bar m}}\right)}, \\
&B= \frac{\sqrt{\pi}}{2}\frac{\sqrt{GM_{\rm bin}}}{\sigma}\frac{\rho_g}{\rho_\star}\frac{m_1}{\bar m}I\left(\frac{\sigma}{c_s}\right)\xi(q)
\end{align}

\noindent
where $W$ is the Lambert W-function. 
We considered the relative velocity in eqs. \ref{eq:a_crit},\ref{eq:rho_crit} to be the velocity dispersion of stars.

The critical density for which a given semimajor axis $a$, which is considered soft when taking into account only stellar hardening, becomes hard due to the contribution from gas hardening, is given by 

\begin{align}\label{eq:rhogcrit}
\rho_{\rm{g,crit}}= \frac{2}{\sqrt{\pi}} \rho_\star \frac{\ln \Lambda_{\rm bin}}{I(\sigma/c_s)\xi(q)} \frac{\bar m}{m_1}\frac{\sigma}{\sqrt{GM_{\rm bin}/a}}
\end{align}

\begin{table}
    \centering
    \begin{tabular}{|c|c|c|}
    \hline
        name & notation & fiducial value \\
        \hline
          gas density & $\rho_g$ & $10^5 \ M_\odot \ \rm{pc}^{-3}$ \\
         stellar density & $\rho_\star$ & $10^4 \ M_\odot \ \rm{pc}^{-3}$\\
         velocity dispersion & $\sigma$ & $10 \ \rm{km/sec}$\\
         sound speed & $c_s$ &  $10 \ \rm{km/sec}$ \\
         gas lifetime & $\tau_{\rm gas}$ & $50 \ \rm{Myr}$\\
                  \hline
    \end{tabular}
    \caption{The fiducial values used along the paper unless stated otherwise. }
    \label{tab:fiducial values}
\end{table}

\begin{figure}[H]
    \centering
    \includegraphics[width=1.1\linewidth]{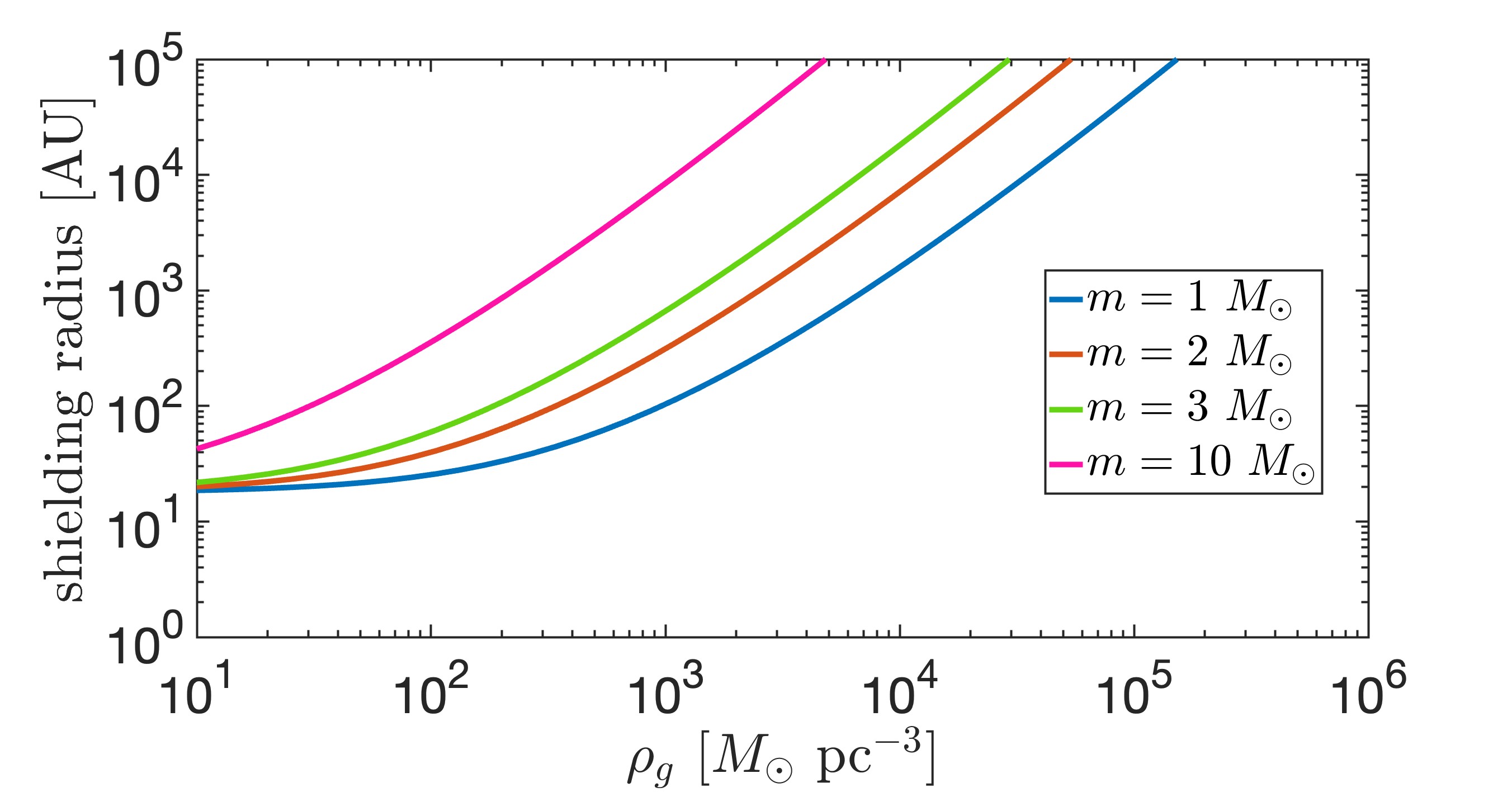}
    \caption{The shielding radius -- the maximal radius for which a binary is considered hard when accounting for the effect of both gas hardening and stellar encounters, as a function of different gas densities, for different binaries of equal masses $m=m_1=m_2$. }
    \label{fig:shielding-radius}
\end{figure}

In Fig. \ref{fig:shielding-radius}, we present the shielding radius, i.e. the largest semimajor axis from which binaries will harden when taking into consideration both stellar hardening/softening and gas hardening, as a function of the gas density. As can be seen, higher gas densities dictate larger shielding radii, as expected, since then the effect of gas hardening is stronger. Higher masses give rise to larger shielding radii, as gas hardening in the regime of soft binaries scales as a positive power of the mass. 

\begin{figure}[H]
    \centering
    \includegraphics[width=1.1\linewidth]{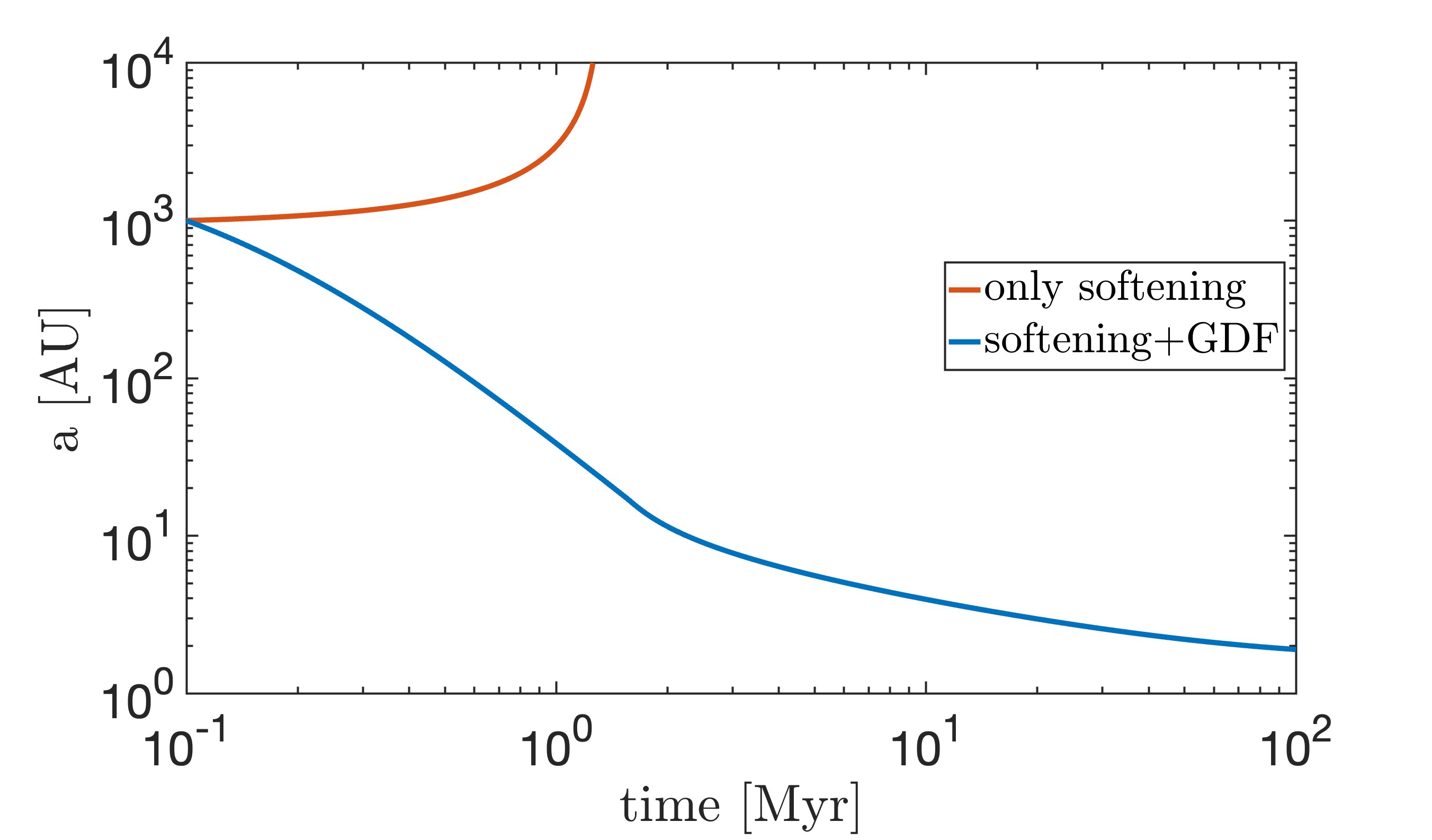}
    \caption{Comparison between the semimajor axes evolution with and without the contribution of gas-shielding, for a binary with $m=m_1=m_2=1 \ M_\odot$.}
    \label{fig:together_compare}
\end{figure}

In Fig. \ref{fig:together_compare}, we examine the semimajor axis evolution of a binary with masses $m=m_1=m_2=1 \ M_\odot$ and initial semimajor axis of $a_0=10^3 \ \rm{AU}$, with and without the contribution of gas hardening. As can be seen, such a binary in a gas-free environment will be considered a soft binary, i.e. its semimajor axis will grow, until finally disrupted. In a gas-rich environment, this binary is essentially hard. The semimajor axis decreases with time, until it reaches a final semimajor axis, as the gas decays and GDF is not efficient anymore.

\begin{figure}[H]
    \centering
    \includegraphics[width=1.1\linewidth]{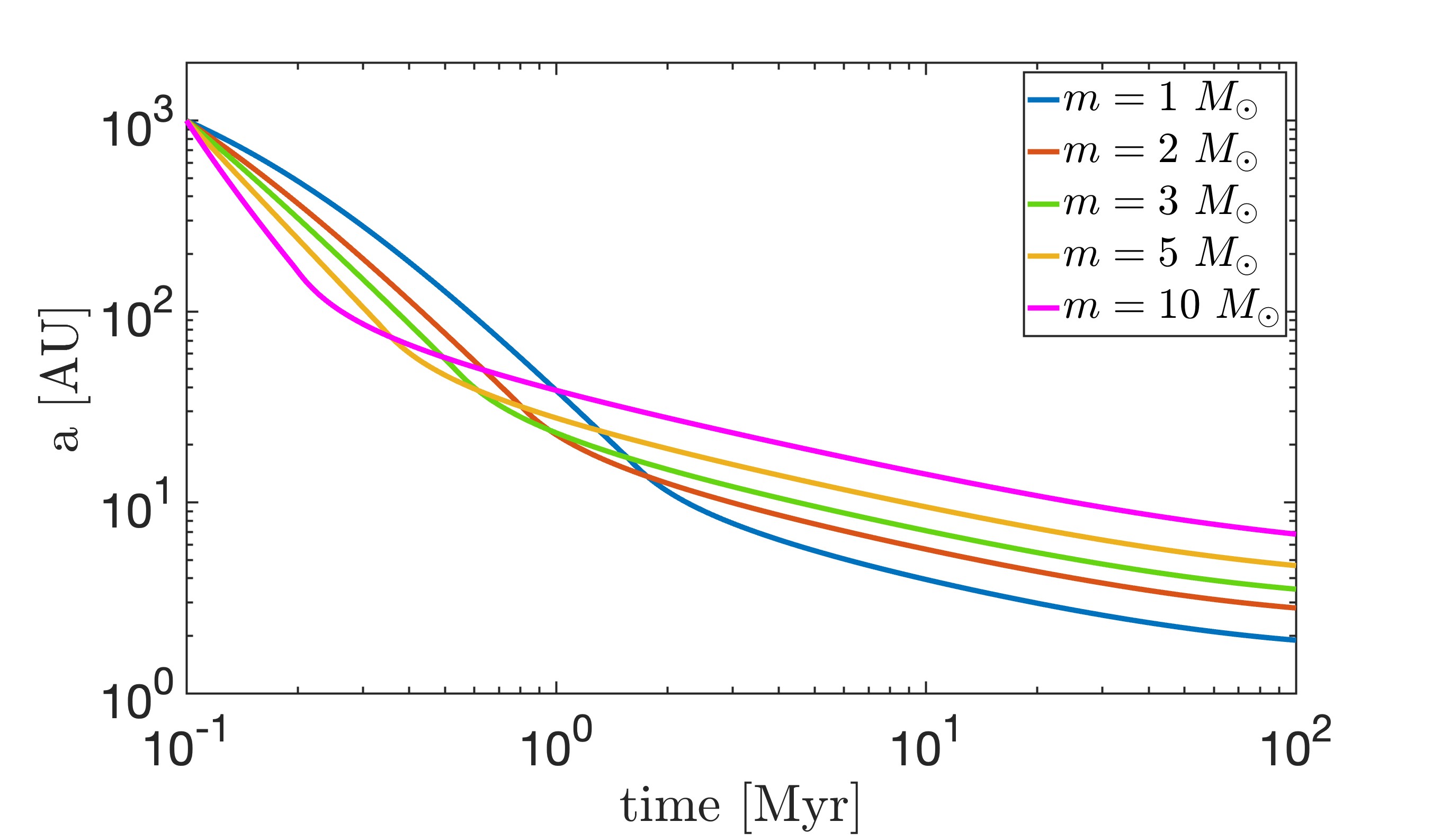}
    \caption{The evolution of binaries with different masses $m=m_1=m_2$, under the effect of gas hardening and hardening/softening.}
    \label{fig:masses}
\end{figure}

In Fig. \ref{fig:masses}, we present the semimajor axis evolution for different masses. The mass dependence is dictated by stellar softening, and scales as $M_{\rm bin}^{-1}$ 
, for hardening there is a weak dependence on the mass scales as $(m_1+m_2+\bar m)/(m_1+m_2)$ 
and for gas-hardening it scales as $m^{-1/2}$, assuming the relative velocity scales as the Keplerian one and as $\sqrt{m}$ when the relative velocity is taken as the velocity dispersion. 
For initially large separations, in the regime where gas hardening dominates, higher masses harden more efficiently. 
Then, at some point, the binary hardens to the hard-soft binary separation, in parallel to the gas decay, and from this point stellar-hardening dominates.

\begin{figure}[H]
    \centering
    \includegraphics[width=1.1\linewidth]{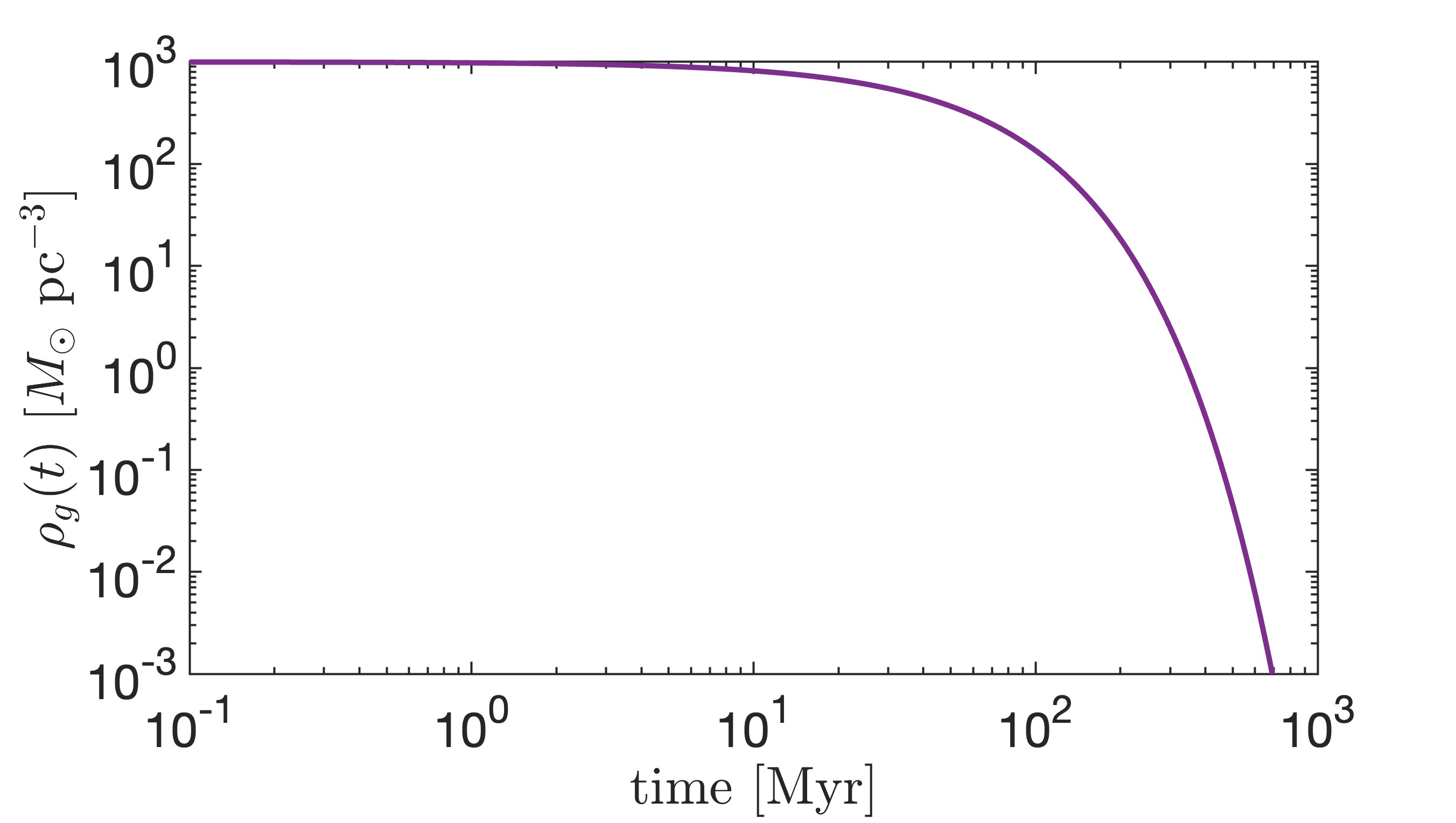}
    \includegraphics[width=1.1\linewidth]{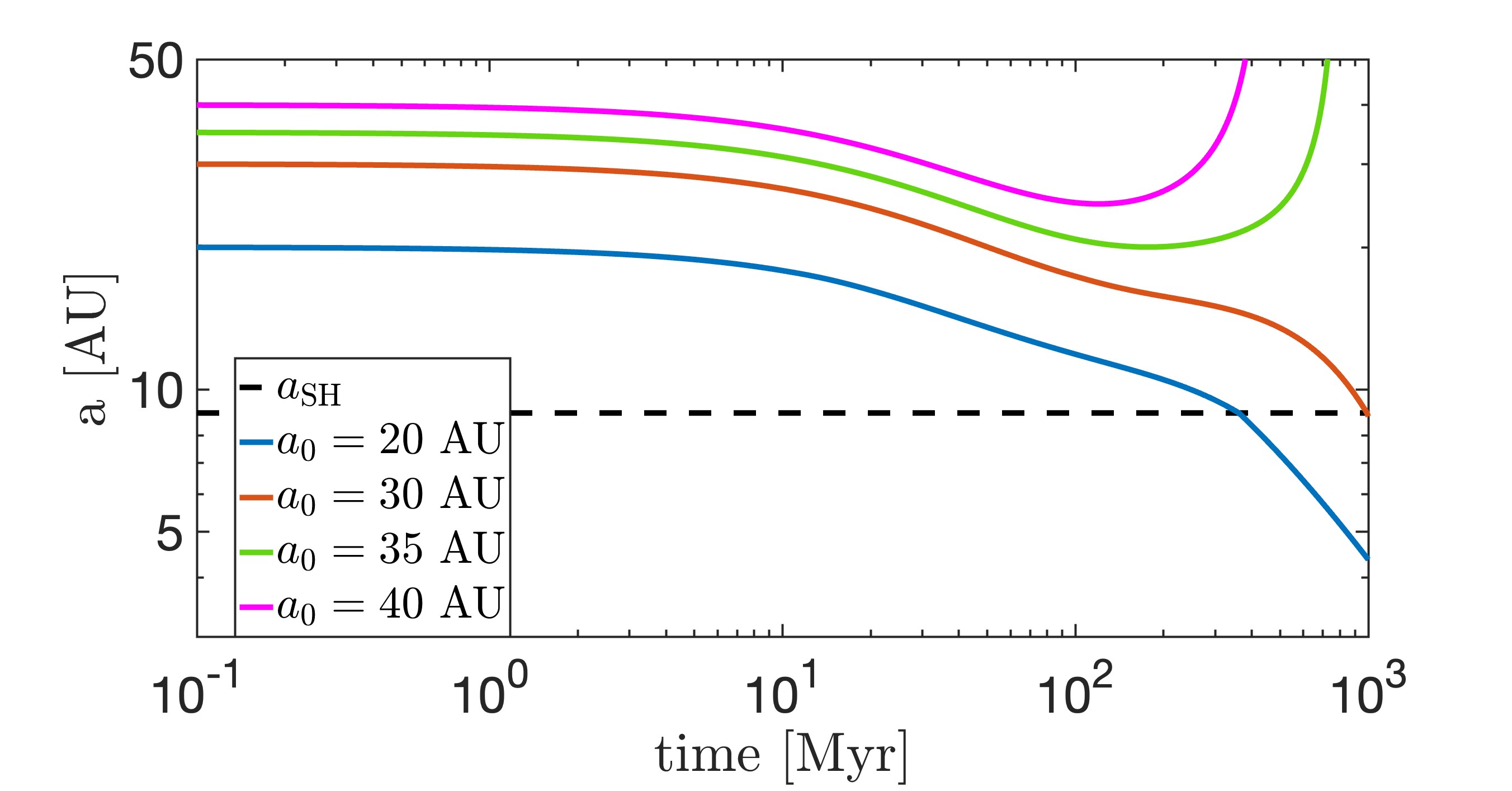}
    \caption{Upper panel: the gas density evolution over time, for initial gas density of $\rho_{g,0}=10^3 \ M_\odot \ \rm{pc}^{-3}$.
    Lower panel: 
    The evolution of different initial semimajor axes for a background density with $\rho_{g,0}=10^3 \ M_\odot \ \rm{pc}^{-3}$, for a binary with masses $m=m_1=m_2=1 \ M_\odot$. The dashed line corresponds to the soft-hard limit boundary as derived based on stellar interactions only.}
    \label{fig:rho_1e3_m1_different_a}
\end{figure}

In Fig. \ref{fig:rho_1e3_m1_different_a}, we present in solid lines the semimajor axis evolution of binaries with masses $m=m_1=m_2=1 \ M_\odot$ and initial background density of $\rho_{g,0}=10^3 \ M_\odot \ \rm{pc}^{-3}$. The dashed line corresponds to the soft-hard limit boundary as derived based on stellar interactions only. As can be seen, the qualitative behavior is different for different initial conditions. While for small semimajor axes, the semimajor axis is monotonically decreasing, for larger semimajor which are above the soft-hard limit, we see that after some decrement, the gas depletes to the point that softening by stellar encounters dominates, and the binary will eventually still be soft, even after the gas-hardening stage. 

To assess the effect of gas hardening on the soft/hard binary population, we carried out a Monte-Carlo simulation, sampling binaries with initial semimajor axes from a log-normal distribution \citep{MoeDiStefano2017} between $1 \ \rm{AU}$ to $200 \ \rm{AU}$, and letting them dynamically evolve.
We then examined the semimajor axes distribution of the binaries at different times.

\begin{figure}[H]
    \centering
    \includegraphics[width=1.1\linewidth]{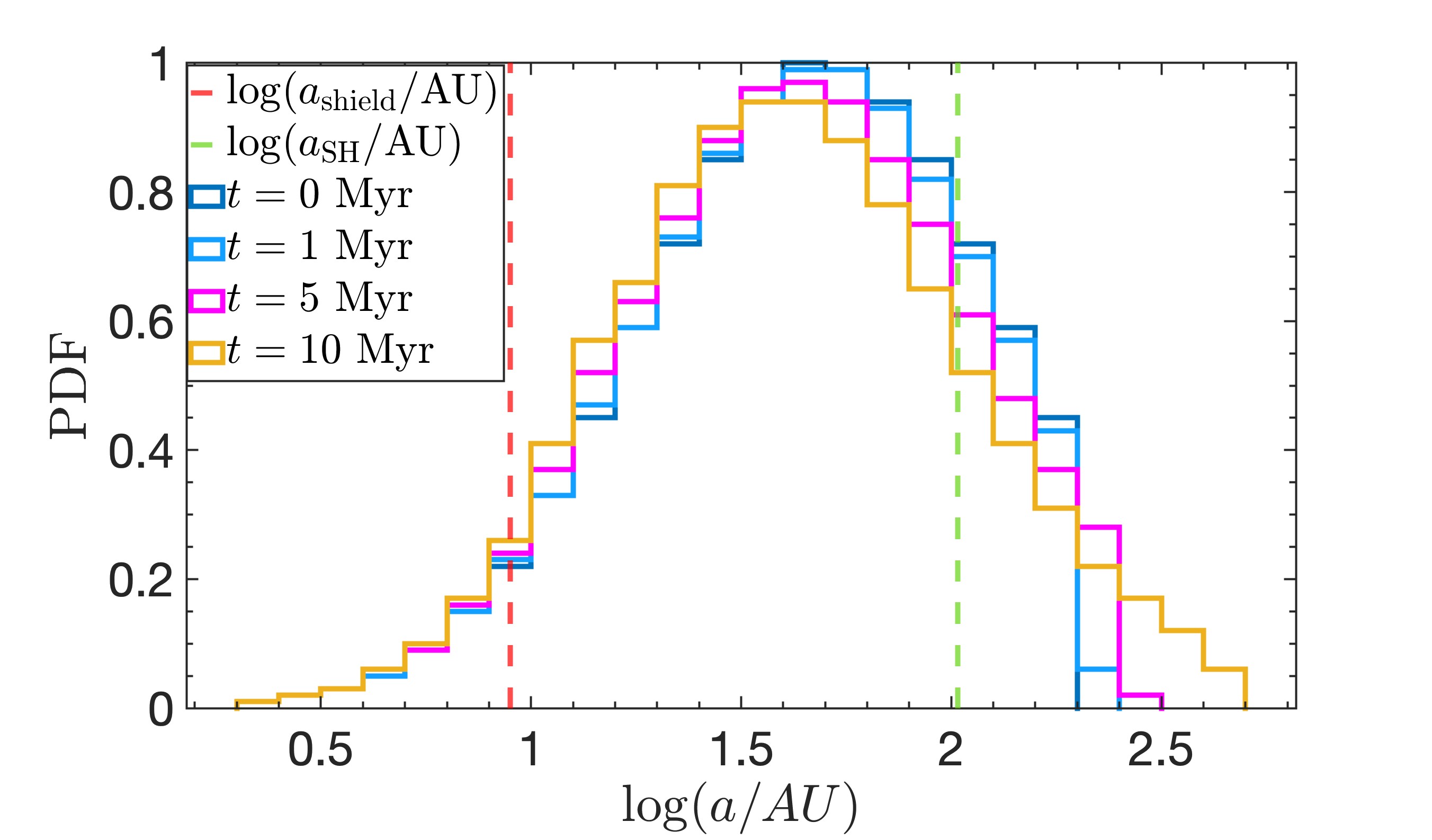}
       \includegraphics[width=1.1\linewidth]{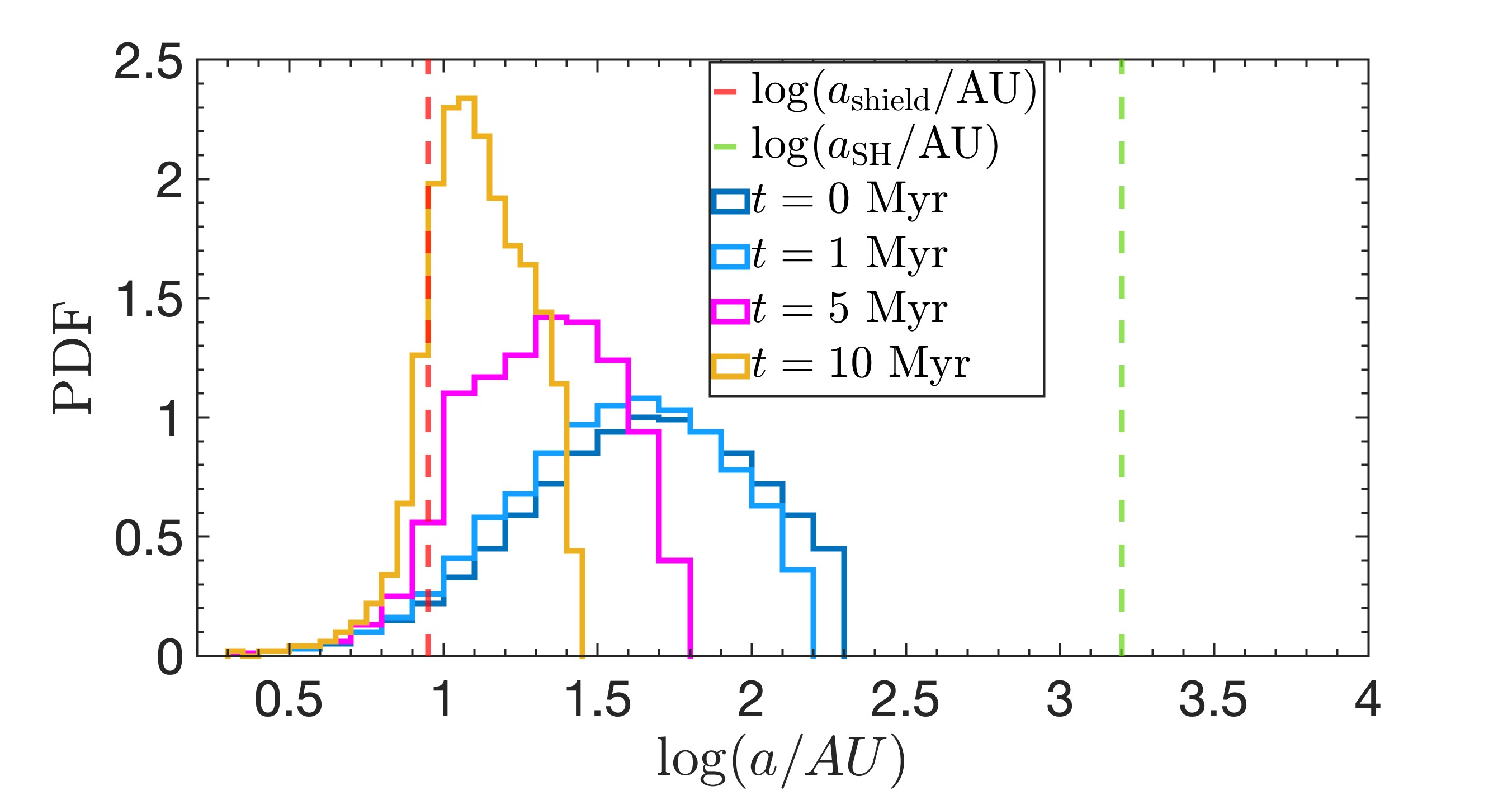}
    \caption{The results of a Monte Carlo simulation, with two equal masses $m=m_1=m_2=1 \ M_\odot$ averaged over $1000$ iterations. The red dashed line corresponds to the critical semimajor axis between soft and hard binaries when considering stellar interactions only (without the effect of gas), and the green dashed line corresponds to the shielding radius as calculated relative to the \textit{initial} gas density. The different solid lines correspond to different times. Upper panel: Initial background density of $\rho_{g,0}=10^3 \ M_\odot \ \rm{pc}^{-3}$. Lower panel: $\rho_{g,0}=10^4 \ M_\odot \ \rm{pc}^{-3}$. Note the different scales. 
    }
    \label{fig:Monte_Carlo}
\end{figure}

\begin{figure}[H]
    \centering
    \includegraphics[width=1.1\linewidth]{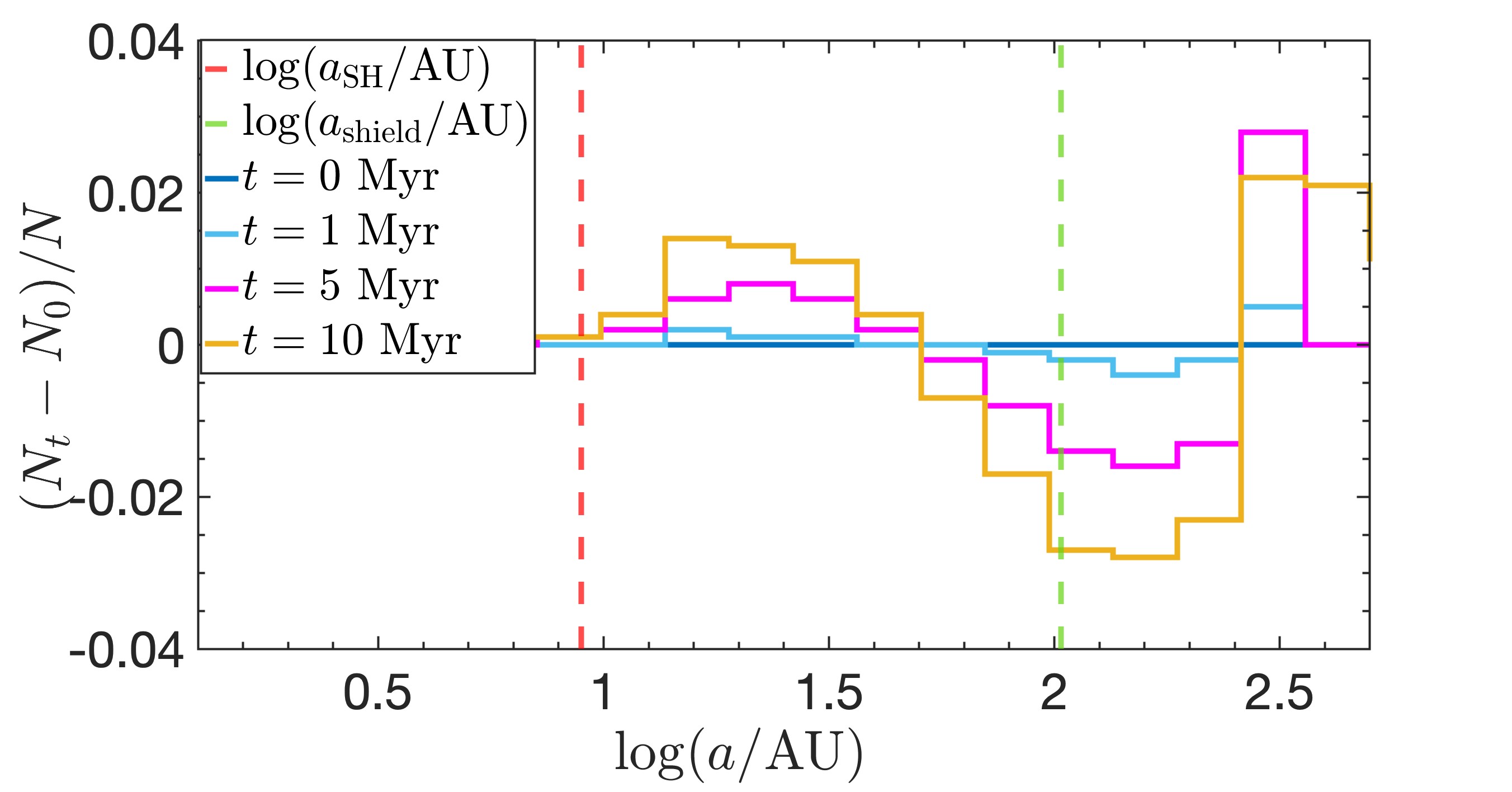}
    \includegraphics[width=1.1\linewidth]{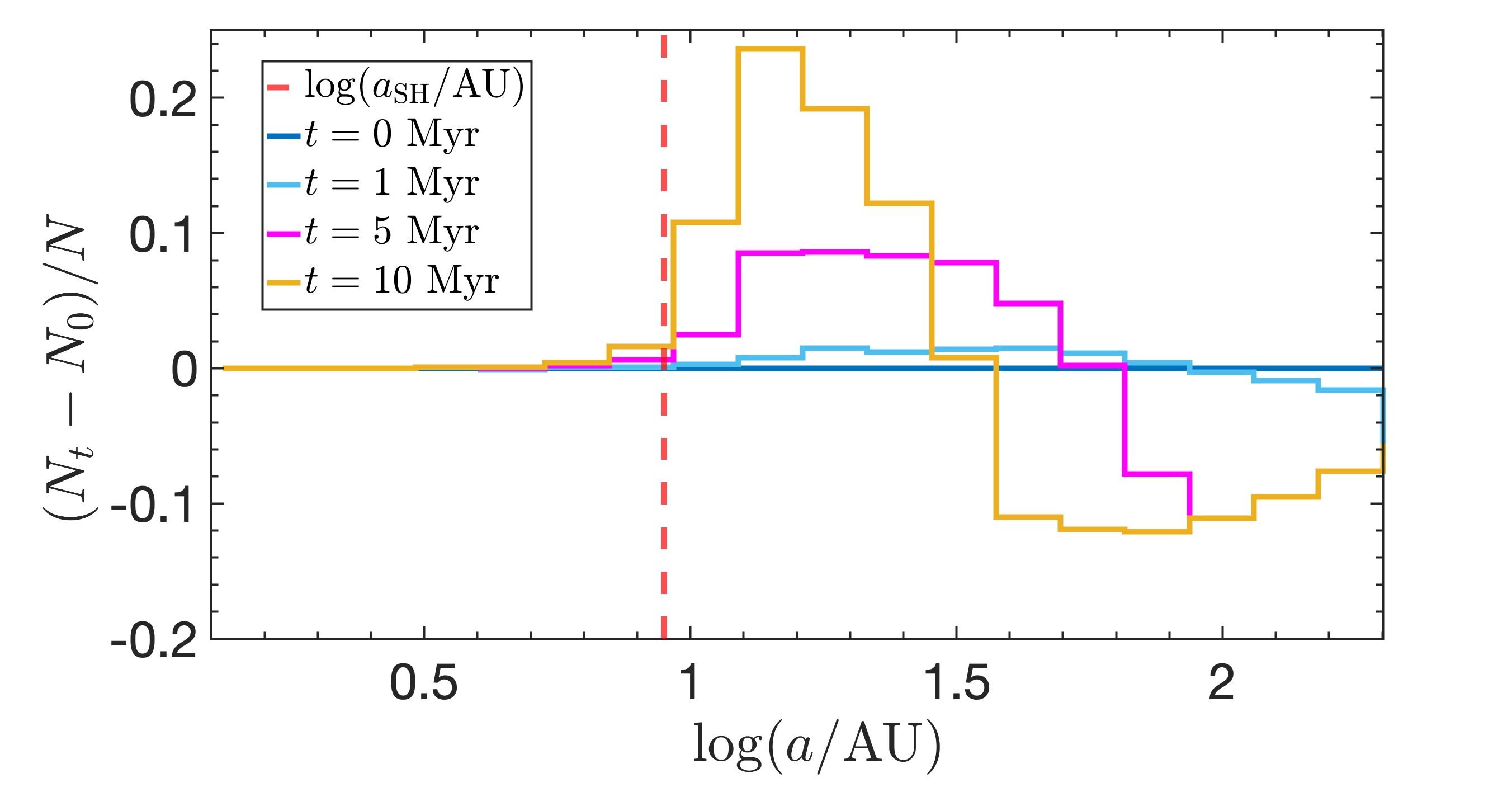}
    \caption{ 
    The normalized difference between the number of binaries in different separations, based on the results of our Monte Carlo simulation, with two equal masses $m=m_1=m_2=1 \ M_\odot$ averaged over $1000$ iterations. The red dashed line corresponds to the critical semimajor axis between soft and hard binaries when considering stellar interactions only (without the effect of gas), and the green dashed line corresponds to the shielding radius as calculated relative to the \textit{initial} gas density. The different solid lines correspond to different times. Upper panel: Initial background density of $\rho_{g,0}=10^3 \ M_\odot \ \rm{pc}^{-3}$. Lower panel: $\rho_{g,0}=10^4 \ M_\odot \ \rm{pc}^{-3}$. Note the different scales.  
    }
    \label{fig:diff_rho1e3}
\end{figure}

Gas shielding changes the distribution of orbital parameters of binaries, as well as their fractions. In gas-free environments, soft binaries are prone to frequent disruptions, that could be evaded in gas-rich environments, since gas hardening could dominate over the stellar softening, under appropriate conditions, leading to hardening instead of softening, as demonstrated earlier. This process could leave significant signatures on the population of binaries. 
To assess the statistical properties of soft/hard binaries in gas-rich environments, we carry out a Monte-Carlo simulation. 
For low-mass, FGK stars we follow \cite{Raghavan2010,MoeDiStefano2017} findings and consider a log-normal distribution centered around $50 \ \rm{AU}$ with a dispersion of $\sigma = 1=\log(10 AU/AU)$. 

We draw binaries with separations between $1-200 \ \rm{AU}$, equal mass binaries with $m=m_1=m_2=1 \ M_\odot$ and initial gas background density of $\rho_{g,0}=10^3 \ M_\odot \ \rm{pc}^{-3}$ and $\rho_{g,0}=10^4 \ M_\odot \ \rm{pc}^{-3}$ for the upper and lower subfigures correspondingly.

In Figs. \ref{fig:Monte_Carlo},\ref{fig:diff_rho1e3}, we present the semimajor axis distribution at different times. As expected, binaries with semimajor axes smaller than the shielding radius (which decreases with time, due to gas depletion) tend to decrease their semimajor axes, even if they are termed as soft binaries when considering stellar interactions only. Hence, the initial binary population is redistributed, such that binaries that are considered hard relative to the combined effect of gas and stars harden, and binaries considered soft, soften, and will eventually be disrupted. It should be noted that we didn't consider here replenishment of binaries and aborted the simulation when a separation of the tidal radius was reached. In a more realistic scenario, further binaries could be formed, either by gas-assisted binary formation, wide binary capture, or any other formation mechanism, adding further complications to our model. As expected, increasing the gas density enhances the effect and could lead in extreme cases to converting all the binaries to be hard. In these cases, gas-assisted binary formation which scales with the gas density, is expected to be highly efficient, potentially leading to the refilling of soft binaries population.  

\section{Discussion}\label{sec:discussion}

Gas shielding could lead to several important implications on the dynamics of binaries and their properties as a population. Here we will discuss our results and possible implications.

\subsection{Implications for other gas-rich environments}

Although we described here the effect of gas shielding in gas-rich clusters only, it is a general process that is expected to take place in any other gas-rich environment, e.g. affecting binary stars in AGN disks and star-forming regions, or binary-planetesimals in protoplanetary disks (with the necessary modifications). It should be noted that in gaseous disks, such as AGN or protoplanetary disks, the effect of shearing could modify the results significantly. In protoplanetary disks, dynamical friction could dominate over the effect of gas dynamical friction, depending on the evolutionary stage of the disk. In star-forming regions, the gaseous epoch is expected to be shorter, and gas shielding will apply mainly for stars rather than compact objects, which are likely not to have formed yet. 

Our results could be used also to set constraints on the gas abundances and lifetimes in these environments, as we could derive expected hard/soft binary fractions for a given amount of gas.  

\subsubsection{Star-forming regions }

In the following, we also demonstrate the process for star-forming regions. 
Star formation takes place in gas-rich environments, starting from molecular clouds that later collapse to prestellar cores and finally pre-main sequence stars (see a detailed review in \citealp{BerginTafalla2007}).
The typical gas mass of the clouds is $\sim 10^6 \ M_\odot$, within a radius of $\sim 10 \ \rm{pc}$ , which dictates a typical gas density of $10^3 \ M_\odot \ \rm{pc}^{-3}$. The velocity dispersion (and the sound speed) in these regions is $\sim 2 \ \rm{km/sec}$ and the typical gas lifetime is $\sim 2 \ \rm{Myr}$ \citep{BerginTafalla2007,Goodwin2013}. Assuming we have a similar density of gas and stars, we consider $\rho_\star=10^3 \ M_\odot \ \rm{pc}^{-3}$, and $n_\star=10^3 \ \rm{pc}^{-3}$. 

\begin{figure}[H]
    \centering
    \includegraphics[width=1.1\linewidth]{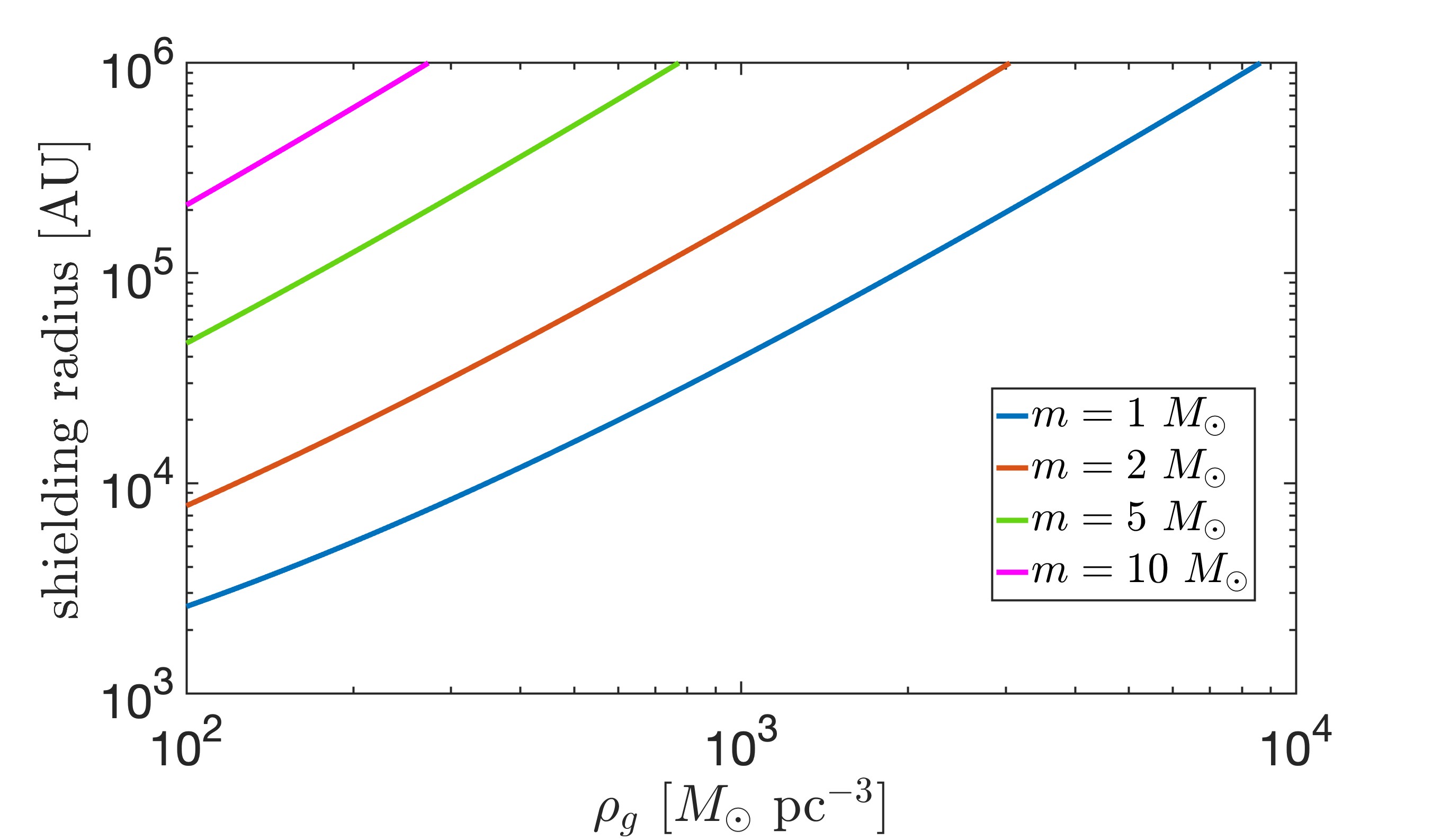}
    \caption{The shielding radius -- the maximal radius for which a binary is considered hard when accounting also for the effect from gas hardening -- as a function of different gas densities, for different binaries of equal masses $m=m_1=m_2$, for the parameters of clouds of star-forming regions.}
    \label{fig:shieldingSF}
\end{figure}

In Fig. \ref{fig:shieldingSF}, we present the shielding radius for the characteristic parameters of star-forming regions. It can be seen that the shielding radii are typically very large, allowing the formed stars to harden instead of soften and get disrupted. Therefore, in these environments, the vast majority if not all the formed binaries are likely to at least partially harden due to a significant contribution from gas hardening.

\subsection{Gravitational waves and other aspects of binary evolution}

Gas shielding naturally leads to an increment in the population of hard binaries, and the survivability of soft binaries. Hence, in gas-rich environments, the occurrence rate of events that require a hard binary could be elevated. In particular, the rates of binary-single and binary-binary encounters giving rise to close approaches, tidal interactions, GW inspirals and collisions will be increased, both due to the larger number of binaries, as well as the existence of more compact binaries; these should affect the rates of GW mergers, thermonuclear explosions from mergers of white dwarfs and other stellar merger, and mass-transfer interacting binaries products. 

\subsection{Hard binary formation}

Most of the three-body encounters that dynamically form binaries produce wide soft binaries that quickly soften and disrupt (\citealp{AarsethHeggie1976,Atalla2024,GinatPerets2024}). Wide binaries could form also by captures \citep{Kouwenhoven2010,wide_binaries}. 
In particular, modeling of soft binaries' steady-state formation and disruption suggests the existence of the order of a single soft binary in a cluster at any given time. However, this steady-state estimate does not account for the existence of gas. Once the gas is included soft binaries that form dynamically can be gas-shielded and become hard-binaries. In other words, the dynamical formation of binaries which is typically inefficient could become a major production form of binaries.
The typical formation rate of hard binaries via gas shielding could be estimated by

\begin{align}
    &\Gamma_{\rm hard}(a)\sim\frac{N_{\rm soft(a)}}{\tau_{\rm shield}}\approx \\ 
    \nonumber
    &\approx 0.12 \rm{Myr}^{-1} N_{\rm soft}(a)\left(\frac{m}{M_\odot}\right)^{1/2}\left(\frac{a}{10^3 \rm{AU}}\right)^{1/2}\left(\frac{\rho_g}{10^4 M_\odot \rm{pc}^{-3}}\right)
\end{align}

 where $N_{\rm soft}$ is the number of formed soft binaries and $\tau_{\rm shield}$ is the gas shielding timescale, given by $\tau_{\rm shield}=|a/\dot a_{\rm GDF}|$, where we considered two equal masses $m=m_1=m_2=1 \ M_\odot$, and since the binary is wide $v_{\rm rel}=\sigma=10 \ \rm{km/sec}$.

In addition, gas-assisted binary capture \citep{Tagawa2020,capture2023,Rowan2023,Whitehead2023,DeLaurentiis2023} could take place in gas-rich environments, initially leading to the formation of soft binaries that will possibly refill the parameter space of hardened soft binaries. 

\subsection{Binary distributions following the gas dispersal}

The gas-rich epoch of the second-generation star formation is truncated by the explosion of supernovae, that clear the cluster from gas. After that, the distribution achieved from the combined stellar-gas dynamics enters a freezout phase, from which another, different distribution of soft binaries could be obtained. In the absence of gas, and under the assumption of thermal equilibrium we derived analytically the separation distribution of soft binaries in clusters and showed that it obeys a powerlaw rule \citep{wide_binaries}. However, after the gaseous phase, even if the distribution of binaries finally reaches equilibrium, the total number of hard binaries will grow, and some soft binaries could be shielded from disruption and remain soft by the dissipation of gas, while others will become hard.
Moreover, there could be global effects in the cluster, such as enhanced mass segregation, as all the stars could be thought of as in a binary with the cluster potential, leading in turn to enhanced stellar scattering and further stellar interactions. 

While we explored some of the effects of gas shielding on the distribution of soft binaries, the long-term effects of producing different binary populations than typically assumed should be further explored by a detailed population synthesis study, accounting for these effects. 

\subsection{Caveats \& future directions}\label{sec:caveats}

\begin{itemize}
    \item Here we modeled the motion in gas using GDF. Qualitatively similar results should be obtained, in principle, using other models, substituting a different gas hardening law in our energy dissipation calculation. 
    \item The gas density profile of a globular cluster during the second-generation star formation is currently unknown. We chose a typical density that corresponds to the total mass of second-generation stars enclosed by a typical volume of the core times an order of unity efficiency factor, but the actual density could vary. 
    
    \item We considered only circular binaries. However, in a more realistic calculation, one should consider also eccentric binaries, for which gas hardening is even more efficient, assuming a flat gas density profile (e.g. \citealp{RoznerGW2023}). 
    
    \item We considered an initial log-normal distribution for the binary separation, which corresponds to the observed current distribution of FGK binaries \citep{MoeDiStefano2017}. However, in the early stages of the cluster, there could be a significant contribution from binaries that were captured via gas-assisted binary capture, with a preference towards large separations comparable to the Hill radius \citep{capture2023}. These binaries will enrich the soft binary population.
    In addition, the distribution of more massive stars tends towards a log-uniform distribution, and would therefore give rise to initially harder binaries, on average. Stellar evolution, not considered here would also affect the distribution at late times; in particular compact-object binaries would not follow the field star binary distributions.
    
    \item We ignored gas accretion on the binaries. As the mass will increase due to accretion, the total effect of gas hardening should be strengthened when considering this effect as well. 
\end{itemize}

\section{Summary}\label{sec:summary}

In this paper, we discussed the effect of gas hardening on soft/hard binaries and studied both particular examples and the total effect on the population. We defined the shielding radius of a binary as the largest separation in which a binary is hard relative to the joint contribution of stellar interaction and interaction with the gas. We showed this radius could be significantly larger than the 'standard' soft-hard boundary when considering stellar interactions only, and hence the effect of gas-hardening is significant and could revise the dynamics of binaries in gas-rich environments, as we demonstrated for globular clusters during their second-generation star formation.  

While the interaction of binaries with gas has been extensively studied during the last few years, mainly in the context of AGN disks, there are still various unexplored directions. The dynamics of populations in gas are essentially different than the ones in gas-free regions, and the distributions of binary populations change accordingly. 

Finally, we focused on binary shielding in gas-rich GCs, but the phenomenon applies in general for every gas-rich environment and hence should leave important signatures on the binary population in these environments, such as AGN disks and star-forming regions. 

\section*{Acknowledgements}

We thank Barry Ginat and Evgeni Grishin for fruitful discussions.  
MR acknowledges the generous support of Azrieli fellowship.
MR and HBP acknowledge support from the European Union's Horizon 2020 research and innovation program under grant agreement No 865932-ERC-SNeX.




\bibliographystyle{aasjournal}

\bibliography{example1.bib}





\end{document}